\documentclass[%
reprint,
superscriptaddress,
showpacs,
 amsmath,amssymb,
 aps,
pre
floatfix,
]{revtex4-1}
\usepackage[english]{babel}
\usepackage{graphicx}
\begin{document}
\title{A novel entropy recurrence quantification analysis}


\author{G. Corso}
\affiliation{Departamento de Biof\'isica e Farmacologia, Universidade Federal do Rio Grande do Norte, Natal, RN, Brazil}
\author{T. L. Prado}
\affiliation{Laborat\'orio Associado de Computação e Matemática Aplicada, Instituto Nacional de Pesquisas Espaciais, S\~ao Jos\'e dos Campos, SP, Brazil}
\affiliation{Instituto de Engenharia, Ci\^encia e Tecnologia, Universidade Federal dos Vales do Jequitinhonha e Mucuri, Jana\'uba, MG, Brazil.}
\author{G. Z. dos S. Lima}
\affiliation{Departamento de Biof\'isica e Farmacologia, Universidade Federal do Rio Grande do Norte, Natal, RN, Brazil}
\affiliation{Escola de Ci\^encias e Tecnologia, Universidade Federal do Rio Grande do Norte, Natal, RN, Brazil.}
\author{S. R. Lopes}
\email{lopes@fisica.ufpr.br}
\affiliation{Departamento de F\'isica, Universidade Federal do Paran\'a, Curitiba, PR, Brazil.}

\begin{abstract}
The growing study of time series, especially those related to nonlinear systems, has challenged the methodologies to characterize and classify dynamical structures of a signal. Here we conceive a new diagnostic tool for time series based on the concept of information entropy, in which the probabilities are associated to microstates defined from the recurrence phase space. Recurrence properties can properly be studied using recurrence plots, a methodology based on binary matrices where trajectories in phase space of dynamical systems are evaluated against other embedded trajectory. Our novel entropy methodology has several advantages compared to the traditional recurrence entropy  defined in the literature, namely, the  correct evaluation of the chaoticity level of the signal, the weak dependence on parameters, correct evaluation of periodic time series properties and more sensitivity to noise level of time series. Furthermore, the new entropy quantifier developed in this manuscript also fixes inconsistent results of  the traditional recurrence entropy concept, reproducing classical results with novel insights.
\end{abstract}
\maketitle

\section{Introduction}

The development of nonlinear dynamics has influenced so many areas of science, from physics, chemistry and engineering to  life science and ecology. From economics to linguistics, and more recently neuroscience \cite{ecology,
economics,linguistics,strogatz_2014, izhikevich_2007,neuroscience}. In last decades, the increasing mathematical knowledge of the complex structures of nonlinear systems has provided successful tools to the understanding of  irregular space and temporal behaviors displayed by collected data in all applied sciences. 
Time series analysis has turned to be a key issue providing the most direct link between nonlinear dynamics and the real world  \cite{kantz_2003}. 

In many cases,  time series data are not extracted from  linear systems. In these cases, linear data analyses bring unsatisfied results  
since linear time series analysis methods are characterized by averaged quantities like mean values or variances, as well as (auto)-correlations, not capturing  the nonlinear aspects of the signal  \cite{kantz_2003}. 

On the other hand, nonlinear analysis tries to extract information from the underlying dynamics of the data. In this way, nonlinear techniques supply new tools for data diagnostics using a whole set of quantities, such as divergence rates, predictability, scaling exponents and entropies in symbolic representation. All those methods are based on more general phase space properties \cite{kantz_1998} such as recurrence properties and others. 

Nowadays, nonlinear time series are a central issue in science. 
An important characteristics of such time series is the presence of natural and non trivial periodicities, characterized by repeating segments of the signal in a rather complex way \cite{gustavo_2017}. In the last decades several sophisticated mathematical tools have been  developed to characterize these periodicities. We cite, for instance, the development of wavelet analysis that have enable scientists to explore in detail the time-frequency structure of signals \cite{Mallat}. Another technique designed to analyze the statistical periodicities in several scales was the detrended fluctuation analysis, a contemporary development of Hurst analysis \cite{peng,dfa2,Gustavo_PlosOne:2014,GZSL-PRE:2012}.  

We explore in this work a new quantifier of nonlinear analysis of time series based on properties of phase space recurrences. The modern concept of recurrence dates to Henri Poincar\'e work \cite{poincare} and it is a fundamental attribute of dynamical systems. A modern visualization method known as recurrence plot (\textbf{RP}) was introduced in \cite{eckmann}, and is constructed from the  recurrence  matrix $\mathbf{R}_{ij}$ defined as:
\begin{equation}
\mathbf{R}_{ij}(\epsilon)=\Theta(\epsilon-||\mathbf{x}_i-\mathbf{x}_j||), \mathbf{x}_i \in \mathbb{R}, \,i,j=1,2,\cdots,M,
\label{rec_plot}
\end{equation}
where $\mathbf{x}_i$ and  $\mathbf{x}_j$ represent the dynamical state at time $i$ and $j$, $\Theta$ is the Heaviside function, $M$ is length of the analyzed time series and $\epsilon$ is the threshold or vicinity parameter, consisting  of a maximum distance between two points in a trajectory such that both points can be considered recurrent to each other. 

In this way, the \textbf{RP} is a symmetric matrix of ``ones'' and ``zeros" where an one (zero) intends for recurrent (non recurrent) points in phase space.  The recurrence analysis technique is conceptually simpler than spectral linear and nonlinear analysis like Fourier or Wavelets  and numerically easier to be performed since it does not have to decompose the signal within a basis \cite{Mallat}. Instead, the \textbf{RP} is computed using repetitions (or recurrences) of segments of the signal which produce a time mosaic of the recurrence  signal, an imprinting of signal time patterns. 

Based on the statistical properties of the recurrence plot, a large number of quantifiers have been developed to analyze details of a \textbf{RP} \cite{Marwan:PR2007}. Many of them, deal  with statistical properties such as mean size, maximum size, frequency of occurrence of diagonal, vertical or horizontal recurrence lines. An important class of recurrence quantifiers are those that try to capture the level of complexity of a signal. As an example, we mention the already known entropy based on diagonal lines statistics \cite{Zbilut1992,Webber1994}. This quantity has been correlated with others dynamical quantifiers as, for example, the largest Lyapunov exponent, since both  capture properties of the complexity level of the dynamics \cite{Marwan:PR2007}. 

Nevertheless, at this point, it is important to mention that sometimes the diagonal entropy, (ENTR), as defined in the literature \cite{Marwan:PR2007} behaves in an unexpected way, indeed, a quite known problem occurs in the dynamics of the logistic map, namely ENTR  decreases despite the increase of nonlinearity. In fact, to deal with that, the literature has presented an adaptive method to compute recurrences to conciliate the behavior of a decreasing ENTR with the increasing complexity of the logistic map  \cite{letellier_2006}. 

Here we develop a new entropy recurrence quantifier based on the formal definition of system entropy making use of all microstates displayed in a recurrence plot, not just diagonal or horizontal lines. In order to introduce the new quantifier, we consider a recurrence plot and select on it random samples of microstates, small square matrices of $N \times N$ elements (we show results for matrix sizes up to $4\times4$). Using square matrices samples of the \textbf{RP} it is possible to define an ensemble of microstates. As we shall see, the microstates reflect the dynamical recurrent patterns of the time series.  Indeed,  the definition of an entropy quantifier based on microstates frequency provides a good estimator that 
correctly capture the relation between the entropy quantifier and the level of complexity and/or chaoticity, fixing the behavior displayed by the former diagonal entropy quantifier ENTR. As we will show, making use of just captured data, it also provide good results for all major dynamical changes that occurs in dynamical systems.

The rest of this paper is organized as follows:  in section $2$ we briefly show the recurrence technique and  present our  methodology. In section $3$ we apply the new methodology to three well known time series: the first is a periodic harmonic (sine) signal perturbed by a white noise, the second is the time series of the discrete logistic map and the last one is well known nonlinear flux described by Lorenz equations. Finally in section $4$ we discuss our results and point out future perspectives.

\section{Methodology} 

The methodology section is divided in three parts. First, we introduce the recurrence analysis technique based on the recurrence plot $\mathbf{R}_{ij}$, as defined in Eq. (\ref{rec_plot}). In addition, we show some quantifiers used in the literature to summarize the information of the \textbf{RP} that will be used in the manuscript to promote the adequate comparison with our method. In the next subsection we define a novel way to represent microstates extracted from the recurrence matrix, which is used to calculate our proposed entropy.  Finally, in the last subsection we detail the microstate set used to construct the new proposed entropy. 

\subsection{Recurrence plots and recurrence quantification analysis}

The recurrence plot is a graphical binary representation of the complex recurrence patterns extracted from time series \cite{eckmann} or spacial profiles \cite{didi_2006, thiago_2014}. The recurrence plot was firstly developed by Eckmann {\it et. all.} \cite{eckmann} and further explored by several authors.  A good compilation in literature of this issue  is found in \cite{Marwan:PR2007}. Important characteristics of the recurrence plots are the presence of finite size diagonal lines, indicating periodic signals or recurrence segments and isolated points suggesting stochastic and/or chaotic signals. An accurate extraction of specific features of a time series can be obtained by using a set of tools, developed initially by Zbilut and Webber  \cite{Zbilut1992,Webber1994} as measures of signal complexity based on the recurrence matrix. These tools are called recurrence quantification analysis (\textbf{RQA}) or recurrence quantifiers.

The \textbf{RQA} studies different aspects of the recurrence plot, from the density of recurrent (non recurrent) points to the statistics of vertical (horizontal) or diagonal lines  \cite{Marwan:PR2007}. In order to avoid problems with very large recurrence plots, when analyzing long time series, it is convenient to divide the original time series into smaller sub-series (or windows) with size $K$ such that $K \ll M$, being $M$ the size of the entire trajectory.  For each window, we construct a recurrence matrix that is used to compute the recurrence quantifiers. The simplest \textbf{RQA} is the recurrence rate (RR) defined as the density of recurrent points in $\textbf{R}_{ij}$. 

An important question in recurrence analysis is the measure of diagonal lines that represent recurrence segments of trajectories. Diagonal lines are $\textbf{R}_{ij}$ structures  parallel to the  lines of identity defined as $\mathbf{R}_{i+k,j+k}~=~1 \, (i,j = 1,2,\cdots M-\ell;k = 1,2,\cdots, \ell)$,  $\mathbf{R}_{i,j}=\mathbf{R}_{i+\ell+1,j+\ell+1}=0$,  where $\ell$ is the length of the diagonal line. Two pieces of a trajectory following a diagonal line undergo for a certain time (the length of the diagonal) a similar evolution, once they have visited the same region of phase space at different times. This is the key idea behind recurrence and thus a clearcut signature of a deterministic behavior in the time series. 

Accordingly,  $P(\ell) = \{\ell_i ; i = 1 , 2 , \cdots K\}$  is the frequency distribution of the lengths $\ell$ of diagonal lines. It is also described by $P(\ell)=\sum^{K-\ell-1}_{i,j=1}(1-\textbf{R}_{i,j})(1-\textbf{R}_{i+\ell+1,j+\ell+1})\prod^{\ell}_{k=1}\textbf{R}_{i+k+1,j+k+1}$, for $K$ the maximum length of the diagonal lines. The determinism quantifier is defined by DET$=\sum_{\ell=\ell_{\rm min}}^{\ell_{\rm max}}\ell P(\ell)/\sum_{i,j=1}^K \mathbf{R}_{i,j}$ with $\ell_{min}$ ($\ell_{max}$) the minimal (maximal) diagonal line. It was  reported that the inverse of  $\ell_{max}$ (the so called DIV - divergence quantifier) is related with the largest positive Lyapunov exponent \cite{eckmann,Trulla}. 

The vertical (horizontal) lines in $\textbf{R}_{ij}$ are associated to laminar states, common in intermittent dynamics \cite{Marwan:PR2007}. The laminarity is another common quantifiers based on vertical (horizontal) lines in \textbf{RQAs}, and it is constructed in a quite similar form as the determinism.  In fact,  laminarity  is defined as  LAM$=\sum_{v=v_{\rm min}}^{v_{\rm max}}v P(v)/\sum_{i,j=1}^K \mathbf{R}_{i,j}$ with $v_{min}$ ($v_{max}$) the minimal (maximal) vertical (horizontal) lines. The frequency distribution of vertical (horizontal) lines can be written as $P(v)=\sum^{K}_{i=1}\sum^{K-v-1}_{j=1}(1-\textbf{R}_{i,j}) (1-\textbf{R}_{i,j+v+1})\prod^{v}_{k=1}\textbf{R}_{i,j+k+1}$.

It was reported the use of the distribution of diagonal lines $P(\ell)$ for a different quantifier of recurrences, based on the Shannon entropy \cite{Marwan:PR2007}. The general equation for Shannon entropy is as follows:
\begin{equation}
{\cal S}=-\sum_{i=1}^{Q} \; p(i) \; \textrm{log}\; p(i),
\label{eq:ShannonEntropy}
\end{equation}
where $p(i)$ evaluates the probability of occurrence of a specific state $i$, $Q$ is the number of accessible states, and ${\cal S}$ captures, in this sense, how much information resides on such collection of states. The Shannon entropy is an open tool that can be adapted to any probability space $p(i)$. The recurrence plot space of probability assumes many different possibilities, for instance the distribution of diagonals $p(\ell)=P(\ell)/\sum_{\ell=1}^{K} P({\ell})$. In that context equation (\ref{eq:ShannonEntropy}) 
assumes the form: ENTR$=-\sum_{\ell=\ell_{\rm min}}^{\ell_{\rm max}}p(\ell)\; \textrm{ln}\; p(\ell)$. Despite the use of this tool in the literature \cite{Webber1994}, it  presents  serious problems as reported in \cite{letellier_2006}. While the entropy was primarily conceived as a quantification of disorder, this first approach, based on entropy applied to chaotic systems (\textit{e.g.} Logistic Map), provides a unsatisfactory result, sometimes indicating a more organized regime for an arising chaoticity levels.

\subsection{A new entropy of the Recurrence Plot}

In this paper we developed a novel way to extract information from the recurrence matrix.
To properly define an entropy we introduce a new concept of microstates for a  \textbf{RP} that are associated with features of the dynamics of the time series.  These microstates are evaluated by matrices of dimension $N\times N$ that are sampled from the \textbf{RP}.  The matrices can assume several configurations as can be seen in Fig. \ref{fig2} for the  particular situation  $N=2$. The total number of microstates for a given $N$ is $N^{*}=2^{N^{2}}$. The microstates are populated by $\bar N$ random samples obtained from the recurrence matrix such that $\bar N=\sum_{i=1}^{N^{*}} n_{i}$, where $n_i$ is the number of times that a microstate $i$ is observed.

For $P_i=n_i/\bar N $, the  probability related to the microstate $i$, we define an entropy of the \textbf{RP} associated with the probabilities of occurrence of a microstate as 
\begin{equation}
S(N^{*})=-\sum_{i=1}^{i=N^{*}} P_{i}  \;\textrm{ln}\; P_{i}. 
\label{eq:GilbertoEntropy}
\end{equation}

 A clear advantage of new methodology to compute the entropy using Eq. (\ref{eq:GilbertoEntropy}) over the former diagonal entropy is the possibility of  computation of information over all possible microstates.
In addition, is possible to estimate analytically the maximum value of entropy $S(N^{*})$ corresponding to the case in which all microstates are equally probable. For this case $P_i=1/N^{*}$.
In this situation, we have  $S(N^{*})=\ln N^{*}$. The analogous case the minimum value  of the entropy corresponds to the situation in which all sampling matrices are at the same microstate. In this case $S(N^{*})$ is trivially zero. 
We will show in the next topic that our approach does not suffer the inconveniences of the traditional recurrence entropy   quantifier, ENTR, for example, those explored in \cite{letellier_2006}. 

\subsection{The microstates grammar }

The microstates of the recurrence matrix have close relation with patterns of the time series dynamics. For simplicity we analyze in detail the case  $N=2$. 

\begin{figure}[htb]
\includegraphics[width=\columnwidth]{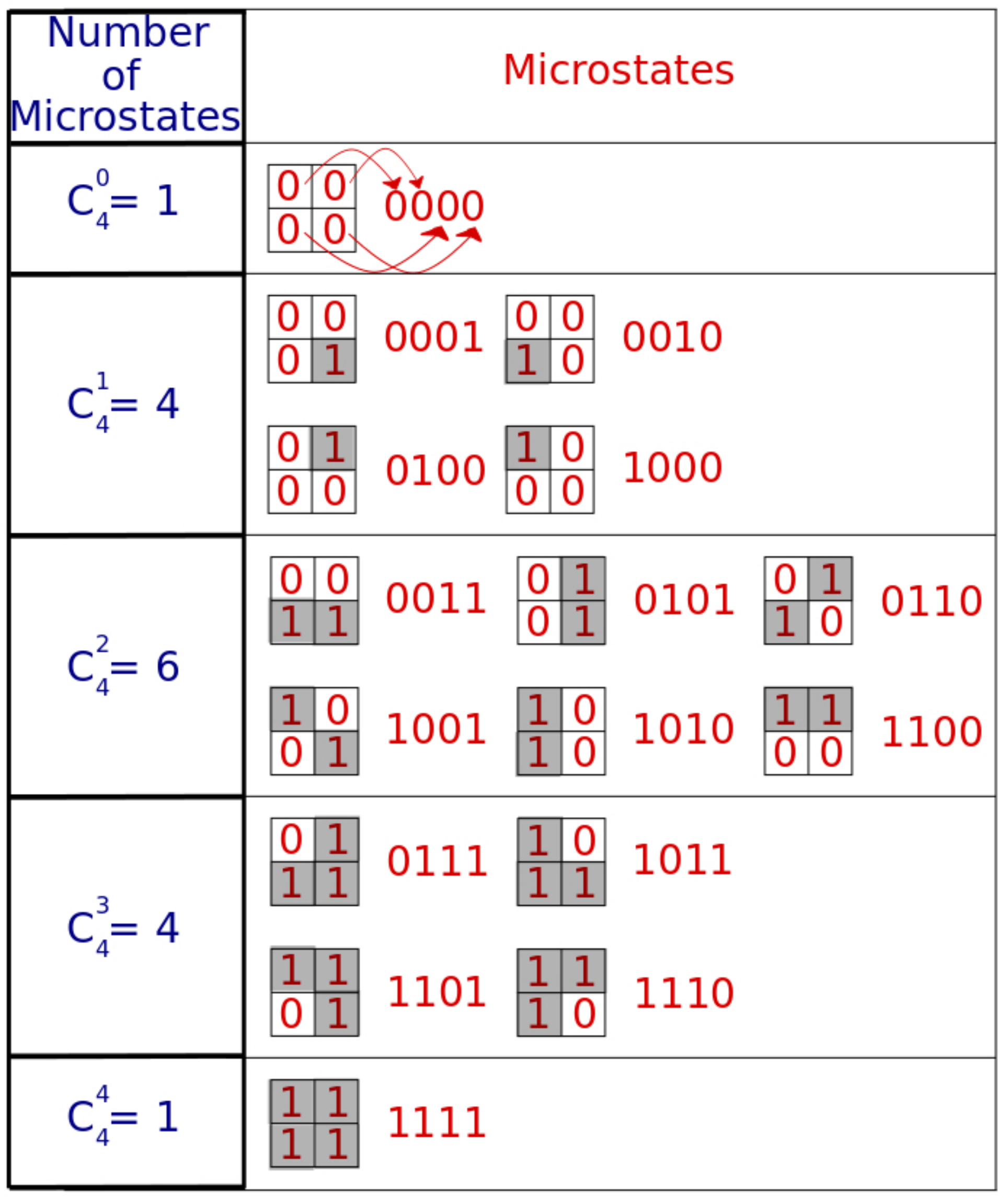}
\caption{\label{fig2} (Color online) All possible microstates of  $N \times N, N=2$ matrix sample along with the correspondent binary recurrent construction. In this setup exists $16$ different combinations that can be grouped according to the amount of recurrent points.  The microstates classes are designated by $C_4^i$, where $i$ corresponds to the number of  recurrent elements in each microstate.}
\end{figure}

For a $2\times 2$ matrix, we initially compute all possible microstates. In this case the microstates are four elements square matrices and each matrix element can assume two values: zero and one.  We illustrate the possible patterns of this problem in Fig. \ref{fig2} in which we show all matrix configurations. Each line is a class $C_{N^2}^i$, $0 \leq i \leq 4$ corresponding to the matrix occupation. The total  number of patterns of a $N \times N$ matrix, for $N=2$ is $2^4=\sum_{i=0}^{N^2} C_{N^2}^i$. The first and last lines of Fig. \ref{fig2} show the trivial situations with all cells are non-recurrent or fully recurrent respectively. The second line, class $C_{N^2}^1$, corresponds to one occupied cell with $C_4^1 = 4$ distinct configurations. The fourth line, class $C_{N^2}^3$, illustrates the situation with three recurrent cells and also with $C_4^3 = 4$ different patterns. The middle line of figure shows the most common situation, where two cells are recurrent and two are non-recurrent. The combination of possibilities is $C_{N^2}^2 = 6$. The $N^4$ matrix patterns of Fig. \ref{fig2} are the total set of microstates used to sample and compute the entropy of the $\textbf{R}_{ij}$. 

We notice that one of the most explored microstate employed in standard recurrence analysis is the (\textbf{1001}) pattern  associated to diagonal lines. The quantifier LAM is related to horizontal and vertical lines whose code patterns  are  (\textbf{1100}, \textbf{0011}, \textbf{1010} and \textbf{0101}). All these four cases are degenerated microstates since all have the same meaning.  In the following paragraphs we analyze in detail the dynamical information of the microstates.  

Firstly, consider the classes of microstates $C_{N^2}^0$ and $C_{N^2}^4$ in which all points are non-recurrent and recurrent respectively. Both situations are typical microstates obtained using a (unappropriated) very small or large value of vicinity parameter $\epsilon$. Using an appropriate $\epsilon$ they will exist, but hardly will be the most frequent microstate. 

When we analyze the dynamical interpretation of the microstates we realize the presence of degenerated microstates.  All microstates in the class $C_{N^2}^1$  are conceptually the same. These microstates do not exist (ideally) in periodic motion or in fixed points dynamics, although they are frequent in chaotic and stochastic series. 
The microstates associated with the $C_{N^2}^3$ class are not properly degenerated, since all of them have their own dynamical characteristics, but in standard analyses they are less frequent than other microstates, because of their intricate combination of recurrences.

\begin{figure}[hbtp!]
\centering
\includegraphics[width=\columnwidth]{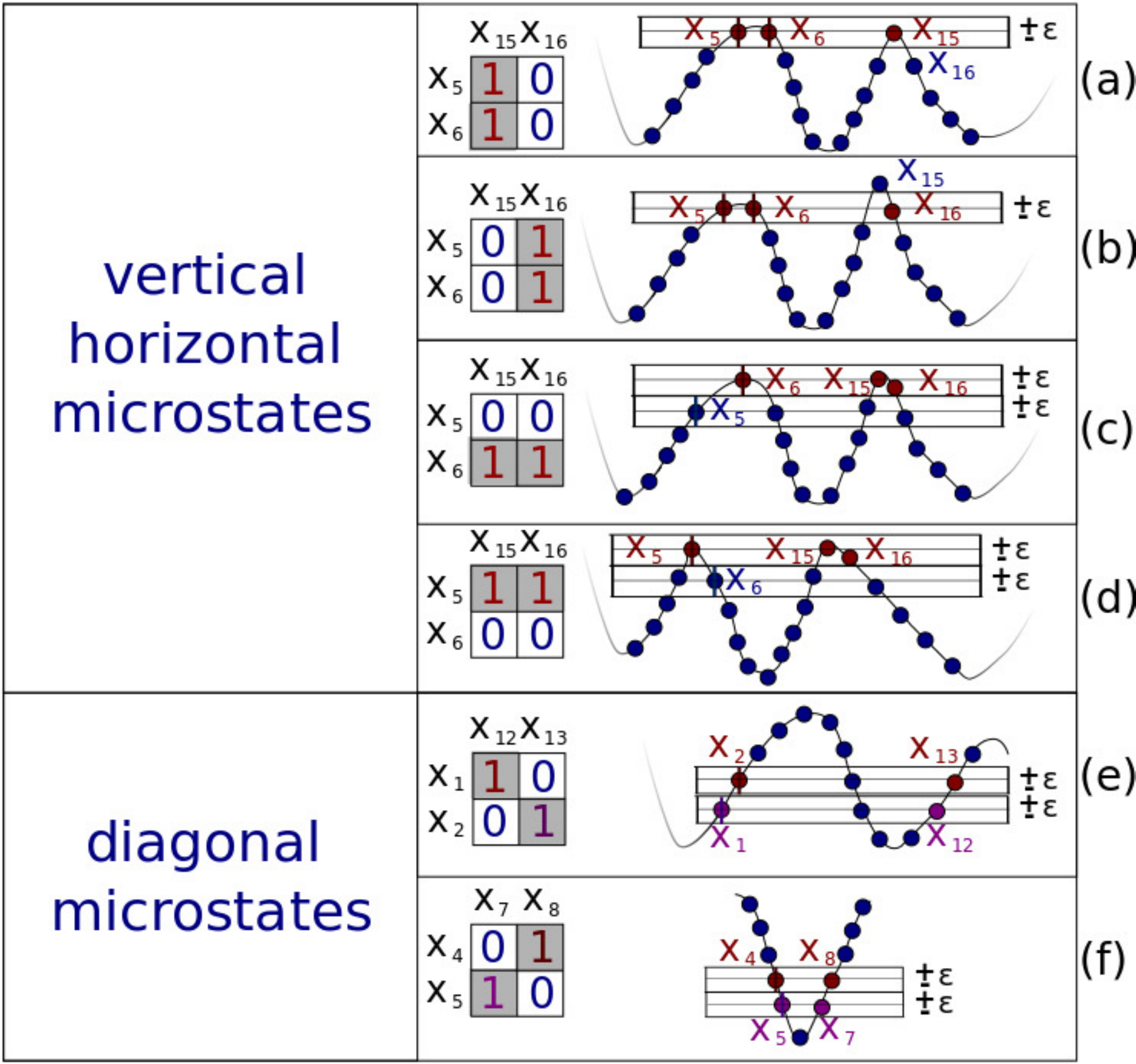}
\caption{(Color online) Scheme of  the structural microstates obtained for $N=2$. Differently from Fig. \ref{fig2} we depict dynamical possible temporal (or spatial) patterns for the case of $N=2$ for vertical, horizontal and diagonal microstates.}
\label{fig3}
\end{figure}

In the recurrence analysis context, the most relevant microstates are all in the  $C_{N^2}^2$  class. For a more accurate analysis, observe Fig. (\ref{fig3}). The group of vertical/horizontal recurrences are related  to laminar states and are characterized by recurrence quantifiers as LAM.
All vertical and horizontal elements have exactly the same information (degenerated microstates), in this case a trajectory recurs to a given position in phase space after some times steps later, and stay nearby for a given amount of time steps. This phenomenon is associated to the symmetry recurrence matrix. 

Inside the class $C_{N^2}^2$, the diagonal microstates are the most important group, examples of temporal  or spatial patterns of theses microstate are observed Fig. \ref{fig3} panels (e) and (f).  Note also that the diagonal microstates are not degenerated. For these microstates there is one that recurs with the same derivative signal, and another one that has a contrary signal of the derivative. Even though, the main information of these microstates is related to recurrent trajectories in phase space that develops nearby, and contain the basic idea of the recurrence quantifier DET.

Using the illustrative case $N=2$, let us consider the dynamics according to the collected data. For periodic signals, just diagonal microstates are observed as shown in  Fig. \ref{fig3}(e). A  chaotic signal shall have a sample of microstates, such as $C_{N^2}^0$, $C_{N^2}^1$ and $C_{N^2}^4$ quite frequent. On the other hand, a smaller portion of diagonal, vertical or horizontal microstates as described in Fig. \ref{fig3} will not be very frequent. Moreover, some residual $C_{N^2}^3$ microstates  are also expected. Finally a stochastic signal should have a more proportional distribution of microstates. 
The most important result in this work regards the evaluation of all these complex behaviors using a comprehensive quantifier: the Shannon entropy.

A similar but more refined and accurate  analysis can be done for $N>2$. Indeed, despite the larger computational effort required to compute the entropy for this case, much more confident results can be obtained. Nevertheless, we have to be aware of the exponential increase of degenerated microstates for larger values of $N$. 

\section{Results}
To explore in details the results obtained by the novel entropy, we apply this tool to a four illustrative data: a white noise, a sine function superposed by white noise, the logistic map signal with and without noise and the Lorenz equations time series. We test the entropy against $\epsilon$,  the vicinity parameter, and the microstate sizes $N$. Moreover, we compare the novel entropy against other well studied recurrence quantification methods. Finally, we discuss the structural advantages of our approach.

\subsection{White noise data}

\begin{figure}[hbtp!]
\centering
\includegraphics[width=\columnwidth]{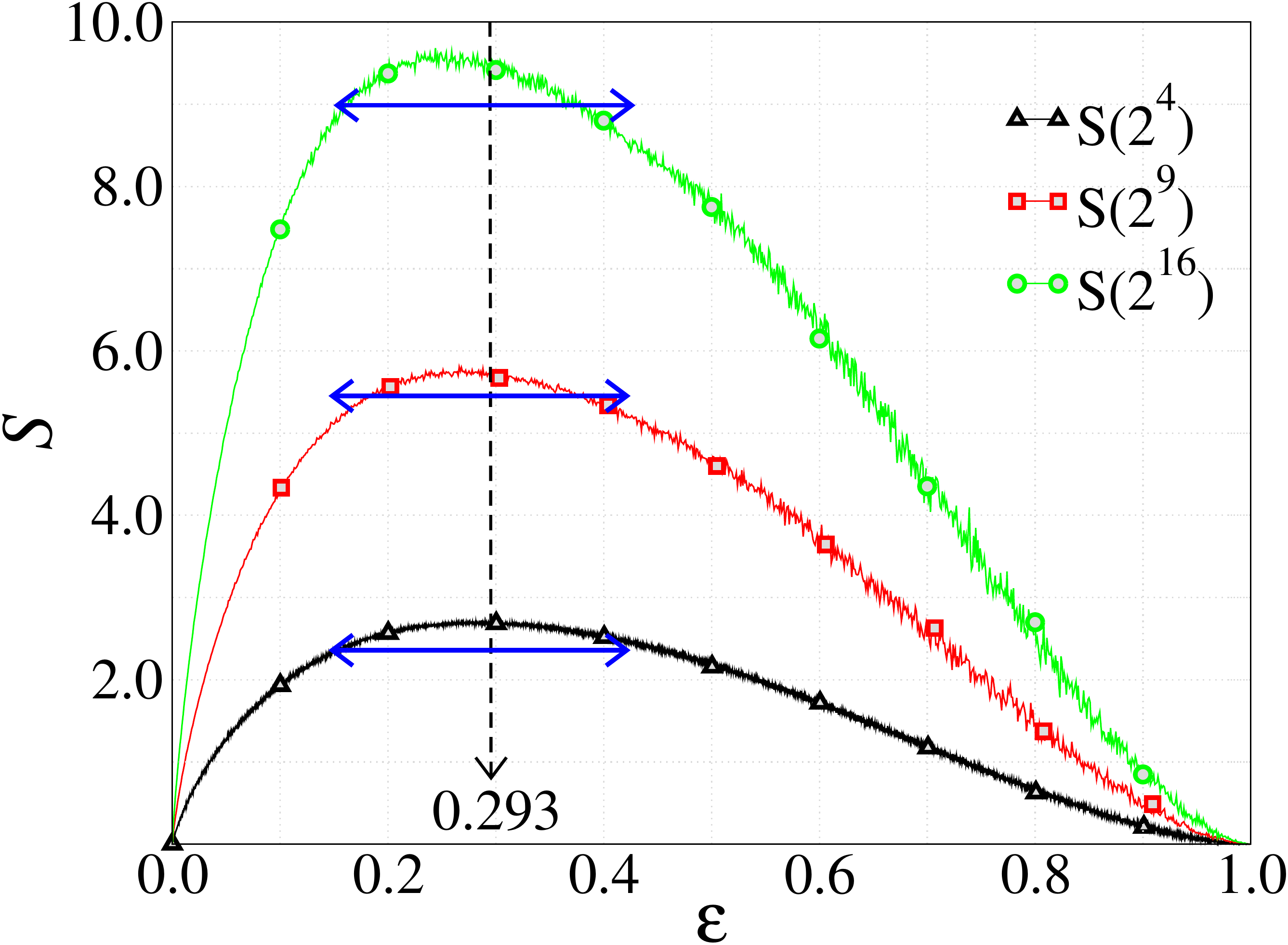}
 \caption{(Color online) Entropy $S$ as a function of the recurrent threshold ($\epsilon$) for 3 different microstate sizes and applied to a white pseudo-random data series. Note that the curve shape is insensible on the amount of possible microstates and that the range of validity of the threshold is quite large, spanning from $0.14  \lesssim \epsilon \lesssim 0.45$. This information stems from the fact that the entropy should reach its maximum for this type of data}
\label{fig6}
\end{figure}

We start our analysis in a random time series with no correlation, namely, a  white noise signal. Fig. \ref{fig6} depicts the entropy $S$ as function of the threshold $\epsilon$.  We employ three values of $N=2,3,4$ as indicated in the legend. 
Observe that, starting in a vanishing value, an increasing  $\epsilon$ leads the entropy to a conceivable interval of validity for $S$. 
This result puts in evidence the resilience of our methodology against this parameter. The cases $\epsilon \rightarrow 0$ and  $\epsilon \rightarrow 1$ should be analyzed in detail. In the limit of small $\epsilon$, there will be just non-recurrent or isolated points in the \textbf{RP}. In this case,  microstates of classes $C_{N^2}^0$, and $C_{N^2}^1$, dominate the distribution and the entropy results to be  small. The opposite regime, for  $\epsilon \rightarrow 1$ the system will present a large number of recurrent points, and the the system will  populate classes $C_{N^2}^3$ and $C_{N^2}^4$, and again, the entropy will result in a minimal value. In this way, intermediary values of $\epsilon$ will produce a richer distribution of microstates among classes and a more confident entropy output. 

We also use the white noise to construct a simple model for maximal entropy in the recurrence plot methodology. The lack of correlation of the white noise implicates in a theoretical maximum entropy. This result is straightforward, but we have to take into account border effects of the recurrence space in the methodology to compute a correct result of $S$. In this simple model is possible to extract the exact value of $\epsilon$ for which the entropy is a maximum. Consider the random data signal $x$ used to compute $S$ in  Fig. \ref{fig6} being $0 \leqslant x \leqslant 1$. It is clear that, for the  vicinity parameter $\epsilon=1$,  RR is maximum and equal to $1$, independently of the point where $\epsilon$ is centered. In this case all points in phase space are recurrent.

When the vicinity parameter is less than $1$, let us say $\epsilon=0.5$ the particular point where it is centered is important, due to the boundaries of the phase space. For example, suppose we are computing the recurrence points of the first value in data. In this case, there is no left recurrence points and a $\epsilon=0.5$ results in just half of the phase space recurrent. In another circumstance, for the same value of $\epsilon=0.5$ but considering the central point of the data, the same methodology  results that, for this particular point, the entire phase space is recurrent.   

\begin{figure}[!htpb]
\includegraphics[width=1.0\columnwidth]{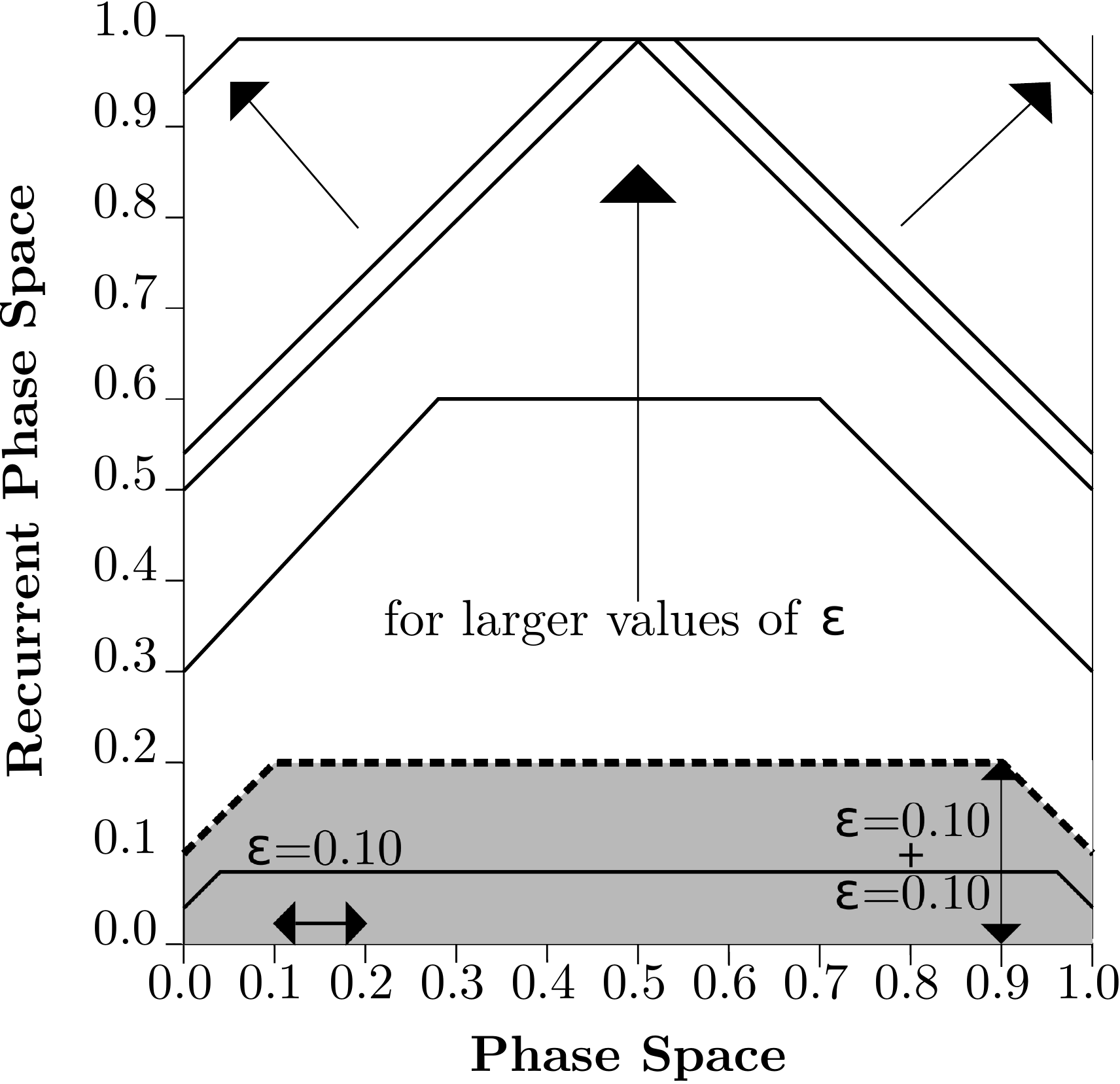}
\caption{Graphical representation of the recurrence rate as trapezoidal areas in the (normalized) recurrence phase space for the white noise signal.  The trapezoidal shape is expected since  points on the left (right) borders of the phase space do not have recurrences on the left (right) sides. Here, the $y$ axis gives us the recurrence percentage of the phase space for different values of $\epsilon$.
}
\label{GeometricalExample}
\end{figure}

In the case of white noise, we expect that the entropy will be maximal for $\mathrm{RR}=0.5$ for which all states are equally populated. However the computation of RR needs to take into account that the $\epsilon$ needs to be considered in all points of the data, even close to the borders where it will capture less recurrent points. We explore in more detail this situation in Fig. \ref{GeometricalExample} where we consider the position where $\epsilon$ is computed in phase space.

Fig. \ref{GeometricalExample} depicts a graphical view of the recurrence space border effect.  The figure illustrates the percentage of phase space recurrence versus the position for what $\epsilon$  is computed for different values of $\epsilon$. We emphasize that this uniformity can be assumed only for a white noise, for simplicity we use a normalized phase space. In this picture RR will be the averaged sampling obtained taking into account the position of $\epsilon$ over all phase space. This estimation is performed  computing the area below each trapezoidal curve as shown. In Fig. \ref{GeometricalExample}, we highlight a particular case for which $\epsilon=0.1$, showing explicitly that it does not correspond to $10\%$ or $20\%$ of the phase space as recurrent, but an intermediate amount.
The general expression for the trapezoidal areas as a function of $\epsilon$ is presented in Eq. \ref{EqNew}. 
Finally, since we expect RR=0.5 for a maximal entropy of the white noise case, Eq. \ref{EqNew} gives us the optimum value of the threshold, namely $\epsilon \approx \mathbf{0.293}$. In fact, despite some numerical  inaccuracy,  Fig. \ref{fig6} confirms this maximum for different values of $N$.
\begin{equation}
\mathrm{RR}=2\epsilon -\epsilon^{2}.
\label{EqNew}
\end{equation}

\subsection{The sine function superposed by white noise}

To further explore the methodology we proceed with a function that continuously change from periodic to stochastic behaviors, depending on a single parameter $p$. We explore the function $y=y(t)$ defined by: 
\begin{equation}	     
y(t) \>  =  \>  \sin(\omega t) + p  \> \mathrm{rand}\,(t)
\label{model1}
\end{equation}
where $p$ is a parameter that controls the random perturbation of the model,  $\mathrm{rand}\,(t)$ is an uniform random function such that $0 < \mathrm{rand}\,(t) <1 $ and  $\omega=0.033$ for convenience. The limit of small $p$ in Eq. (\ref{model1}) makes $y$ a trivial deterministic sine function, whereas for large $p$  the random function dominates and $y$ behaves like a white noise. 
\begin{figure}[hbtp!]
\centering
\includegraphics[width=\columnwidth]{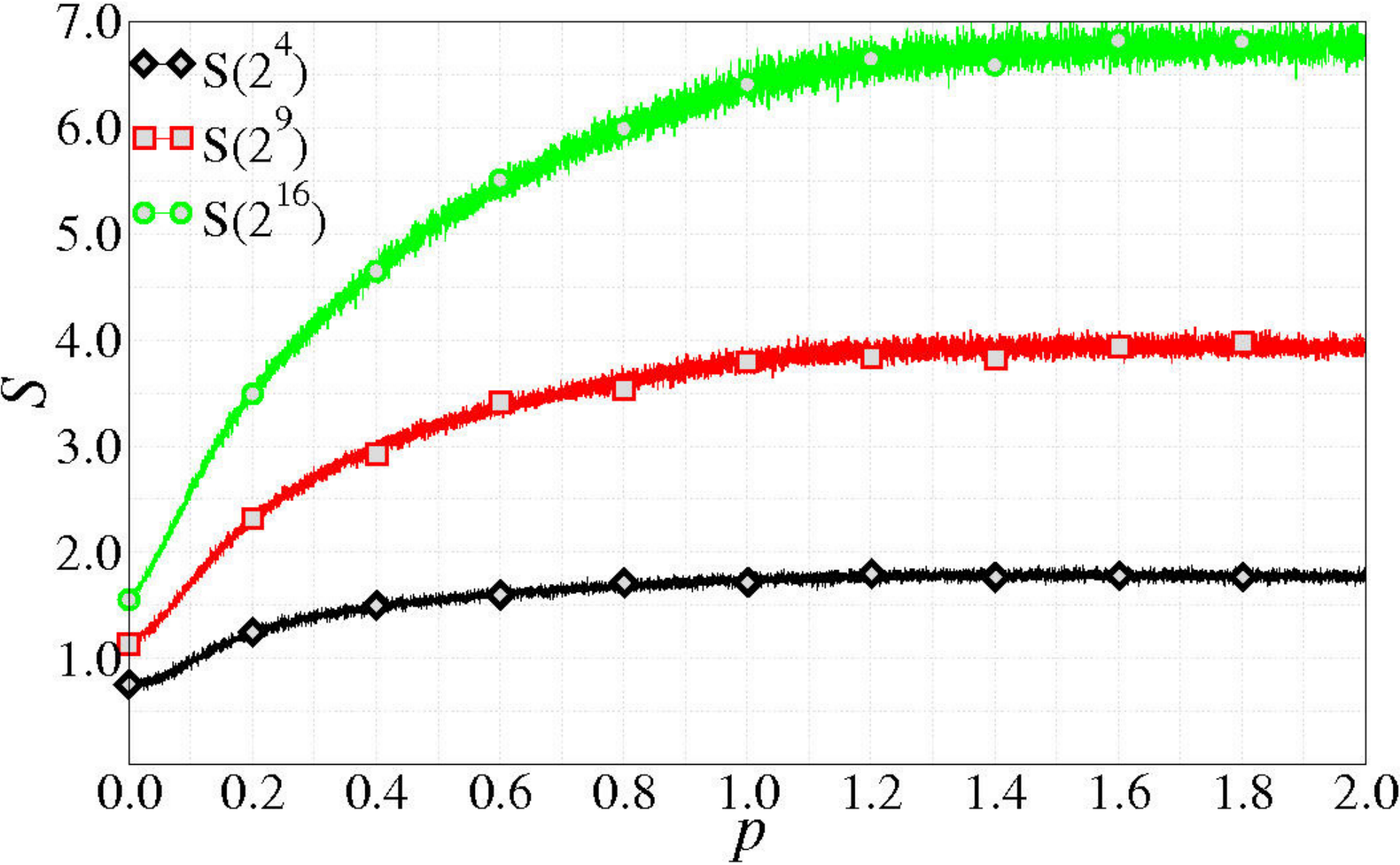}
 \caption{(Color online) Entropy quantifier $S$ as a function of the random parameter $p$  for   $N=2$, $3$ and $4$ for the sine-random signal, Eq. \ref{model1}, using $\epsilon =0.14$. It is noticeable the increase of $S$ as $p$ increases reaching a maximum for large values of $p$ as predicted by Eq. \ref{EqNew}. While the absolute value of entropy changes as a function of $N$,  the general shape of the curve is preserved}
\label{fig4}
\end{figure}

In this context the entropy evaluated from the microstates quantifies the change from an orderly system (\textit{eg.} sine function) to a disorderly system (\textit{eg.} tendency to white noise).  In Fig. \ref{fig4} we show the entropy as a function of $p$ for three microsates lengths,  $N=2$, $3$ and $4$.  We observe in the graphic a continuous and smooth increase of entropy with $p$. Around $p \approx 1.2$ the entropy saturates which characterize the random regime. The opposite situation, $p \rightarrow 0$, shows a minimal entropy regime. For sake of comparison we indicate the theoretical maximal entropy for each $N$:  $S_{max}(N^{*}=2^{4})=2.77$, $S_{max}(N^{*}=2^{9})=6.24$, and $S_{max}(N^{*}=2^{16})=11.09$. We notice in Fig \ref{fig3} that the three curves do not surpass the $S_{max}$  corresponding to the adequate $N$.  
In this way, the $S_{max}$ is a good candidate to a model benchmark that does not depend on the specific analyzed dynamics, but only on the number of microstates. 

In Section II (C) we explored the case of microstates with length  $N=2$, and showed that some microstates are degenerated. When we increase $N$ the number of degenerated states increases dramatically. New microstates appear. As a trivial example, we cite microstates related to larger diagonal lines that represent system states with more   deterministic behavior, since for this case, a trajectory tends to repeat the dynamics of a previous trajectory that visit the same region. The consequence of a larger amount of microstates is the accentuated fluctuation in the entropy, see for instance Figs. \ref{fig6}, \ref{fig4} and  \ref{fig5}.

\subsection{The logistic map}

To explore further our methodology we test the entropy $S$ in a system that presents chaotic regimes. We employ the logistic map \cite{YorkBook,Lichtenberg} defined by the following equation 
\begin{equation}
  x_{n+1} \> = \> r \> x_n (1 - x_n).
\label{log}
\end{equation}
The parameter $r$ controls the non-linearity of the system. The  $r$ is  responsible for the bifurcation cascade route to chaos and windows of periodic behavior that can be seen in the bifurcation diagram Fig. (\ref{fig7})-(f). 
\begin{figure}[hbtp!]
\centering
\includegraphics[width=\columnwidth]{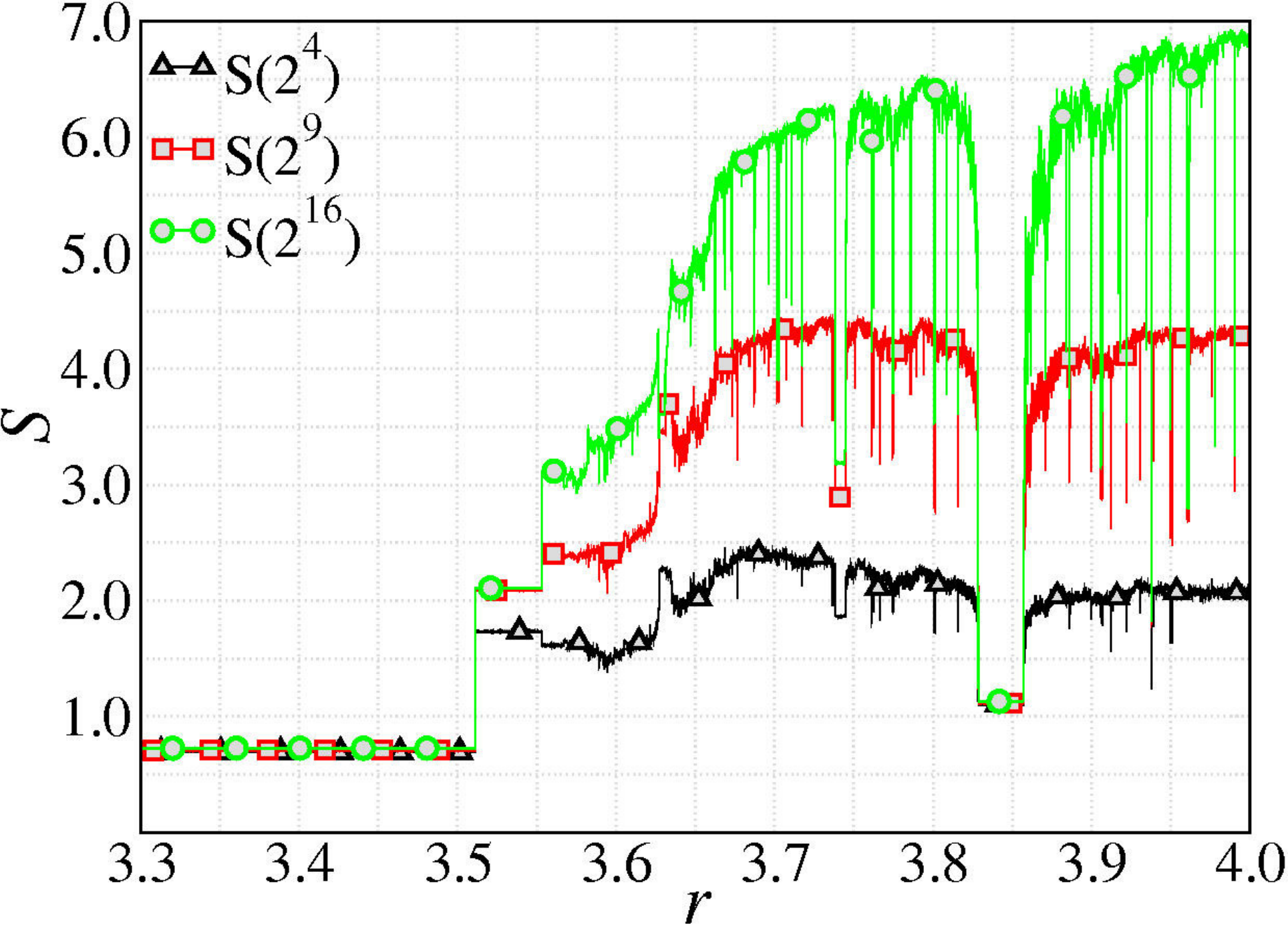}
\caption{(Color online) Entropy quantifier $S$ for the logistic map versus the  parameter $r$. Each curve correspond to a different microstate size $N$. There is a greater resolution on the periodic windows and chaotic regimes for larger $N$. Clearly, $S$ for  $N \geq 3$, ($S(2^{9})$ and $N \geq 4$ ($S(2^{16})$) capture the increase of the complexity of the map dynamics as $r$ increases. Nevertheless, more pronounced results are obtained for $N=4$. All simulations were computed using  $\epsilon=0.14$.}
\label{fig5}
\end{figure}

In Fig. (\ref{fig5}) we show the entropy $S$ as a function of $r$ for $N=2$, $N=3$ and $N=4$. While the general aspect of these curves are similar for all $N$, the resolution  and the main concept of the increasing of the entropy $S$ as $r$ grows is more pronounced for $N=4$. Larger $N$ implies in a larger  number of microstates allowing for a more precise evaluation of the system states. On the other hand, it leads to a larger sampling effort  and  computational time. The main  effect of the $\epsilon$ size can be observed in the doubling period cascading. The entropy jump of value does not occur synchronously with the actual $r$ parameter where a bifurcation occurs. In fact it jumps in a $r$ value a little bigger than the true $r$ value of the bifurcation. It occurs due to the fact that before the observed jump, due to a finite value of $\epsilon$, the quantifier can not distinguish the orbit post bifurcation when compared to the orbit before the bifurcation. It is a common effect of the recurrence analysis. 

To test the advantages of our methodology, we compare our results for $N=4$, $S(2^{16})$  against some  well known recurrence quantifiers. The results are depicted in Fig. (\ref{fig7}) (a)-(e).  To clarify the dynamics, \ref{fig7}(f) displays the bifurcation diagram for the logistic map.  The  quantifiers  RR (panel (c)) and LAM (panel (b)) are almost insensible to the growing complexity of the logistic map as $r$ increases. The quantifier ENTR is extremely erratic and does not transmit the real concept of entropy for the system, since it should reflex the increasing complexity of the map for larger values of $r$. This point, in particular,  was explored in details in the literature and it is pointed  as a drawback of the ENTR quantifier \cite{letellier_2006}.  We should mention also the perturbing fact that ENTR increases in periodic windows and does not show precisely the period doubling cascading window, occurring, for example, for $3.8<r<3.9$ that is associated with enhancing  complexity. In comparison, the entropy $S$ quantifier for $N \geq 4$ ($S(2^{16})$), based on the microstates obtained from the \textbf{RP}, depicts quite well what is going on in the system dynamics. It also reflects correctly the increases of complexity in period doubling cascade. Finally,  $S(2^{16})$ properly capture the increase of complexity of the dynamics as $r$ increases. Furthermore, we should emphasize that $S$ has a really weak dependency of on $\epsilon$, differently of other recurrence quantifiers.
\begin{figure}[hbtp!]
\centering
\includegraphics[width=\columnwidth]{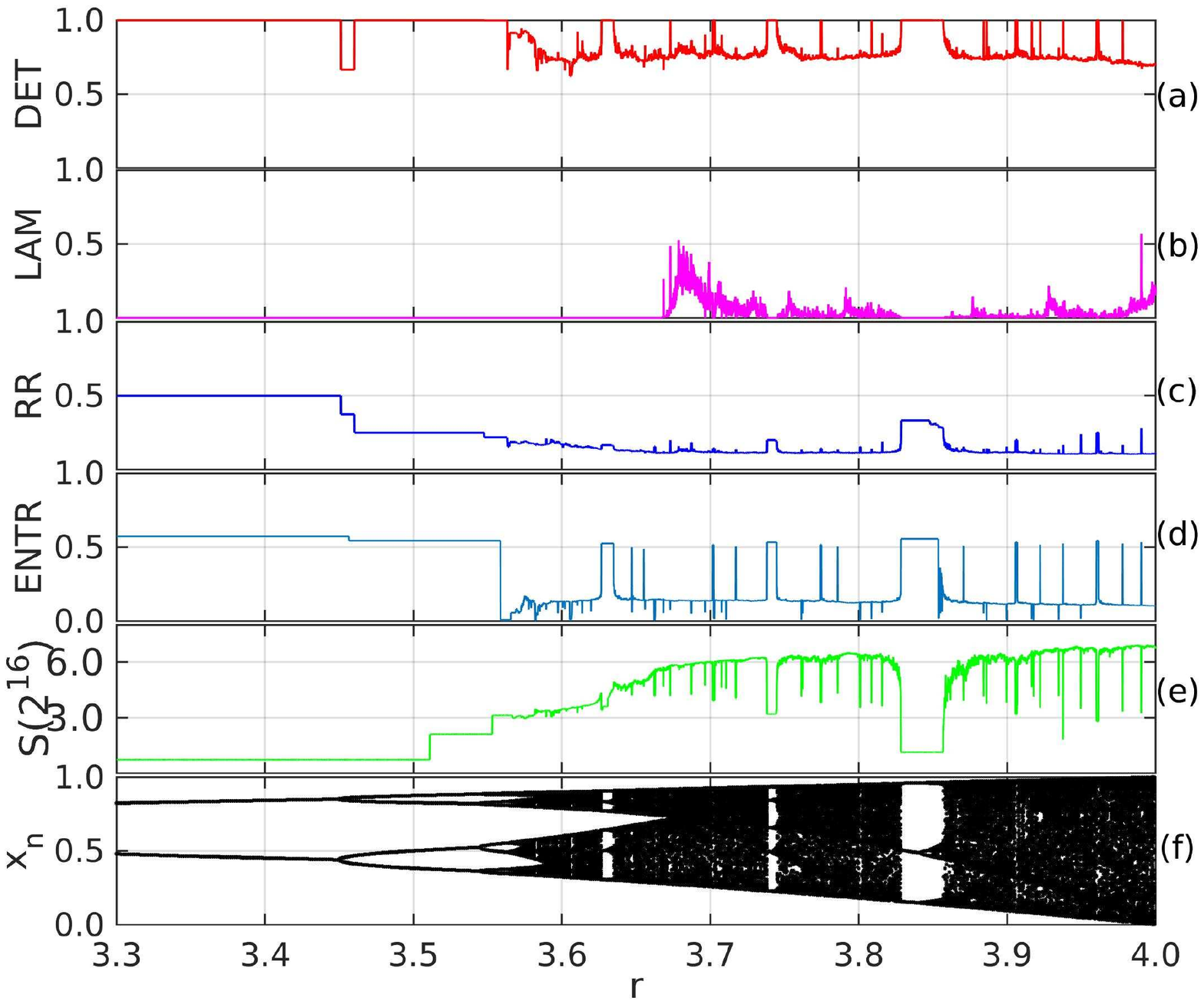}
\caption{ (Color online) Recurrence quantifier analysis obtained from DET, LAM, ENTR, and the new defined recurrence entropy $S$ for $N \geq 4$ ($S(2^{16})$) for the  logistic map, panels (a)-(e). Bifurcation diagram is plotted in panel (f). $S$ was computed using  $\epsilon = 0.14$, $M=1000$ after transient time. Some informations obtained for the new entropy $S$ is similar to the others quantifiers. Nevertheless, note that,  differently from the other quantifiers, $S$ capture the increase of complexity due to the increase of $r$ in chaotic regions.}
\label{fig7}
\end{figure}

\begin{figure}[hbtp!]
\centering
\includegraphics[width=\columnwidth]{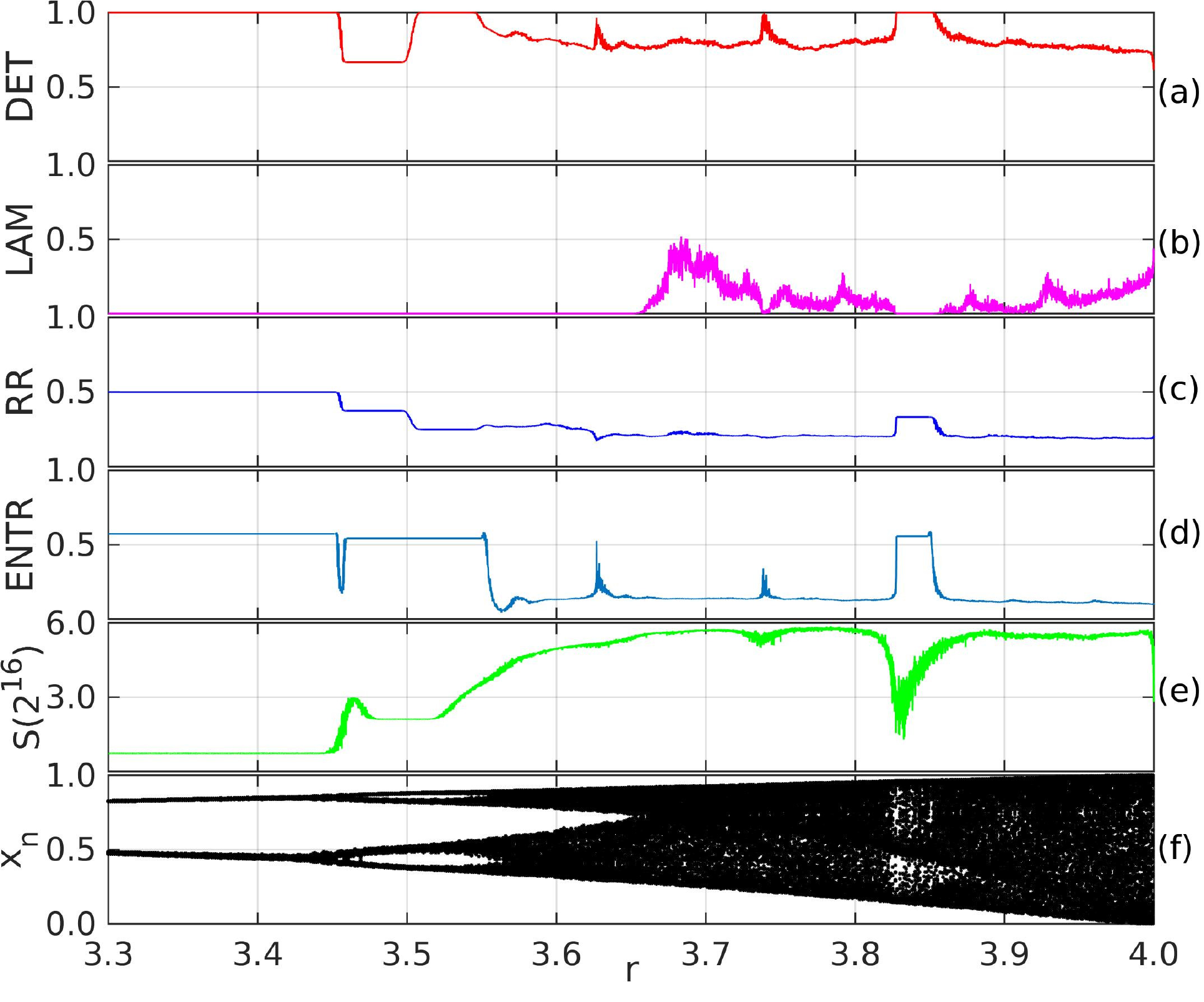}
\caption{(Color online) Recurrence quantifier analysis obtained from DET, LAM, ENTR, and the new defined recurrence entropy $S$ for $N \geq 4$ ($S(2^{16})$) for the  logistic map perturbed by white noise. The noisy bifurcation diagram is plotted in panel (f). In each iteration of the map, white noise perturbations are added in the dynamics of the map, corresponding to $0.5\%$ of the maximum amplitude of the map. $S$ was computed using  $\epsilon = 0.14$, $M=1000$ after a transient time. The new entropy $S$ is quite robust against noise. Different from the other quantifiers, $S$  captures the increase of complexity due to period doubling bifurcations even in the presence of noise as can be observed   in the intervals $3.45<r<3.65$ and $3.8<r<3.9$}
\label{fig8}
\end{figure}

In Fig. \ref{fig8} we explore the logistic map dynamics perturbed by  random noise. In this case the system loses many features that it had for the deterministic case, like the numerous periodic windows within the chaotic region and the perfect periodicity in the well know periodic regions \cite{grassberger}. In this context, all quantifiers roughly brings the same information, but we notice that both, DET and $S(2^{16})$ preserve respectively the reduction/increase in the quantifier while the chaos develop further given the increase of $r$. An important point must be mentioned: Observe that, for the large noisy periodic window, occurring in the interval $3.8<r<3.9$, the entropy $S(2^{16})$ is much more sensitive to changes occurring in the noisy dynamics, when compared to  RR,  DET, LAM  and ENTR. $S(2^{16})$ shows  a clear increase of magnitude in the well known doubling period cascading interval. It is important to state that technically $S(2^{16})$ preserves the same quantitatively and qualitatively advantages against the other quantifiers as have been shown in the case without noise.

\subsection{The Lorenz equations}

In order to illustrate our results for the new methodology to obtain the system entropy applied to a continuous chaotic system. We present results for $S$ using microstate of size $N=4$. The results of the new entropy quantifier are applied to the  classical Lorenz equations \cite{YorkBook}.  The Lorenz model is a well studied continuous dynamical system that present all sort of nonlinearities and even chaos \cite{Lichtenberg,YorkBook}. These equations are a reduction from seven to three differential equations originally developed to model a convection motion in atmosphere \citep{YorkBook}. The three equations that resume the model are written as:
\begin{equation}
     \begin{aligned}
       & \dot{x}=-\sigma (x+y), \\
       & \dot{y}=x(r-z)-y, \\
       & \dot{z}=xy-bz,
     \end{aligned}
     \label{LorenzEquations}
\end{equation}
with three free parameters: the Rayleigh number $r$, the Prandtl number $\sigma$   and the quantity $b$. The system behaves in a periodic way for the set of parameters ($\sigma=10$, $b=8/3$) and $0<r\lesssim 24.06$ \cite{YorkBook}. For $r$ beyond $24.06$ it starts to display chaotic behavior with sporadically appearance of periodic windows. 

\begin{figure}[hbtp!]
\centering
\includegraphics[width=1.0\columnwidth]{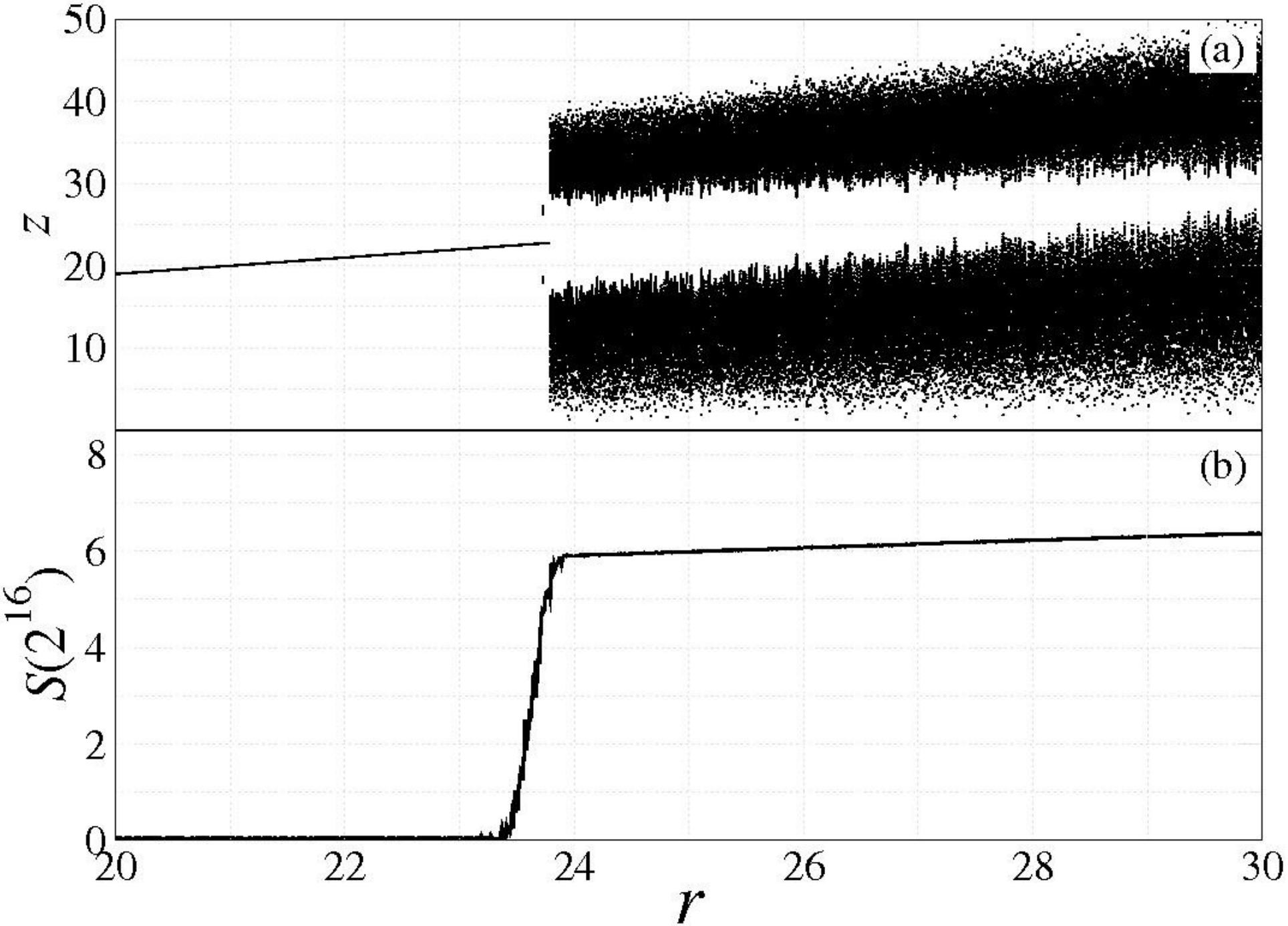}
\caption{The entropy $\textsf{S}$ computed for the Lorenz equations.  We use here $N=4$, $\epsilon = 0.14$, $M=1000$ and  $\bar N =10^{5}$ microstate samples. In (a) we present the bifurcation diagram while in (b) we show the entropy as a function of the parameter $r$ in Eq. \ref{LorenzEquations}. The time series was computed with time step h=$10^{-4}$ and evolved for $10^{9}$ time steps to avoid transient dynamics. As should be expected, the quantifier ${S}$  subtly increases when the chaotic attractor born and, more important, predicts quite well the expanding level of chaoticity of the Lorenz model as the $r$ parameter is variated.}
\label{fig9}
\end{figure}

An example of the application of our methodology to continuous classical chaotic system is depicted in  Fig. (\ref{fig9}). As observed, the transition to the chaotic regime is associated with a dramatic increase in the entropy. Another important aspect is that, after the transition to chaotic states, the entropy display a smooth increase similar to the one depicted in the maximum Lyapunov exponent due to the growing level of system chaoticity. This fact is easily observed by computing the Lyapunov exponent spectrum for the Lorenz system. Note that the maximum entropy for $N=4$ is $S(2^{16})=\ln(2^{16}) \approx 11.09$, but the maximum entropy should exist only in the condition of a random system, putting in evidence a clear difference between the random signal and a deterministic chaotic signal, as expected.

\section{Discussion and Final Remarks} 

This work has explored a new tool to study recurrence patterns in time series. The patterns are evaluated within the framework of Recurrence Plots \textbf{RP}. Our method brings  a novel quantifier that analyzes the microstates obtained from sampled matrices extracted from \textbf{RP}. 
In a broad sense, the manuscript has studied the diversity of the computed microstates  in the \textbf{RP}. 

To quantify the diversity of accessed  microstates we have computed a proper Shannon entropy for the system. To demonstrate the validity of our method, we have applied the concept to a random signal, a simple model composed of a sine signal superposed by  white noise, a discrete chaotic system (the logistic map) and a continuous system exemplified by the Lorenz equations. Moreover, we have tested  the methodology for diverse recurrence vicinity size $\epsilon$ and microstates sizes.
In addition, we have compared our results with  standard recurrence quantifiers indexes: the recurrence rate, the laminarity, the determinism and the traditional entropy of recurrent diagonals as defined in the literature so far.

The main advantage of employing  the Shannon entropy based in microstates, as proposed here, is the fact that it is intrinsic to the meaning of an entropic quantifier. Similar to any other entropies like the Boltzmann entropy, the Kolmogorov-Smirnov entropy, the Shannon entropy, our new methodology to compute the recurrence entropy increases with the complexity of the system. At this point, we should mention that the more traditional diagonal entropy computed as a recurrence quantification does not have this propriety. In fact, this drawback of the quantifier ENTR, has been taken into consideration in the literature \cite{letellier_2006} but, the solution found to solve the problem involve some structural changes in the meaning of recurrence. Our new methodology to compute the entropy does not use any changes in the definition of recurrence. The entropy we introduced in this work naturally increases with signal complexity. 

Another important issue is the computational effort to estimate a quantifier. Using microstates sampling, the amount of data to be analyzed is proportional to the number of possible microstates. For instance when we use $\textsf{N}=3$ a useful sampling can be as low as a few hundreds of microstates. While in other methods, the totality of the $\textbf{RP}$ must be analyzed. In a situation with a standard $10^{3}$ data points, the matrix would reach $10^{6}$ points that should be evaluated to compute a proper quantification of diagonals, verticals or simple density of recurrences.

In addition, it is well known within recurrence plot techniques that the value of vicinity size, $\epsilon$, and also the minimal sizes of diagonal, $l_{\textrm{min}}$, or vertical lines,   $v_{\textrm{min}}$,  affects the value of the quantifiers \cite{eckmann, Marwan:PR2007}. 
The method proposed in this work  shows great robustness against changes of $\epsilon$. The new methodology for the entropy obtains stable results using  microstates with sizes $\textsf{N}\geq 2$ and shows to be adequate to diverse discrete and continuous systems. 

Finally, the methodology here presented remains to be applied to numerous other possible systems, as experimental data, intermittent systems, complex systems dynamics phenomenology, transitions chaos-hiperchaos and many others.

\section*{Acknowledgments}

The authors acknowledge the support of  Conselho Nacional de Desenvolvimento Cient\'ifico e Tecnol\'ogico,  CNPq - Brazil, grant number 302950/2013-3, Coordena\c c\~ao de Aperfei\c coamento de pessoal de N\'{\i}vel Superior, CAPES, trough project numbers 88881.119252/2016-01 and BEX: 11264/13-6 and Fudan\c c\~ao de Amparo \`a Pesquisa do Estado de S\~ao Paulo, FAPESP,  through post-doc scholarship numbered 2015/23487-8.


\end{document}